\documentclass[sn-mathphys]{sn-jnl}
\jyear{2021}

\theoremstyle{thmstyleone}

\theoremstyle{thmstyletwo}%

\theoremstyle{thmstylethree}%

\begin{document}

\title[Gravitational time advancement effect in Bumblebee gravity for EBS]{Gravitational time advancement effect in Bumblebee gravity for Earth bound systems}

\author[1]{\fnm{G.Y.} \sur{Tuleganova}}\email{gulira.tuleganova@yandex.ru}
\equalcont{These authors contributed equally to this work.}

\author[1]{\fnm{R.Kh.} \sur{Karimov}}\email{karimov\_ramis\_92@mail.ru}
\equalcont{These authors contributed equally to this work.}

\author*[1,2]{\fnm{R.N.} \sur{Izmailov}}\email{izmailov.ramil@gmail.com}

\author[3]{\fnm{A.A.} \sur{Potapov}}\email{a.a.potapov@strbsu.ru}
\equalcont{These authors contributed equally to this work.}

\author[4]{\fnm{A.} \sur{Bhadra}}\email{aru\_bhadra@yahoo.com}
\equalcont{These authors contributed equally to this work.}

\author[1,4]{\fnm{K.K.} \sur{Nandi}}\email{kamalnandi1952@yahoo.co.in}
\equalcont{These authors contributed equally to this work.}

\affil*[1]{\orgdiv{Zel'dovich International Center for Astrophysics}, \orgname{M. Akmullah Bashkir State Pedagogical University}, \orgaddress{\street{3A, October Revolution Street}, \city{Ufa}, \postcode{450008}, \state{RB}, \country{Russia}}}

\affil[2]{\orgdiv{Institute of Molecule and Crystal Physics}, \orgname{Ufa Federal Research Centre, Russian Academy of Sciences}, \orgaddress{\street{Prospekt Oktyabrya 151}, \city{Ufa}, \postcode{450075}, \state{RB}, \country{Russia}}}

\affil[3]{\orgdiv{Department of Physics \& Astronomy}, \orgname{Bashkir State University}, \orgaddress{\street{47A, Lenin Street}, \city{Sterlitamak}, \postcode{453103}, \state{RB}, \country{Russia}}}

\affil[4]{\orgdiv{High Energy Cosmic Ray Research Center}, \orgname{University of North Bengal}, \orgaddress{\city{Darjeeling}, \postcode{734 013}, \state{WB}, \country{India}}}

\abstract{This paper is a novel application of the new effect of gravitational time advancement or \textit{negative} time delay, first predicted for static black holes (spin $a=0$), that can be regarded as complementary to the well known effect of positive Shapiro time delay. We shall extend the Shapiro time delay formalism up to third PPN order using the recently proposed spinning ($a\neq 0$) black hole solution of the Lorentz symmetry breaking (LSB) Bumblebee gravity that is believed to reveal signatures of quantum gravity at low energies. Adopting two practical examples of signal propagation along Earth-Moon and Earth-Satellite configurations, we shall calculate the influence of the Bumblebee parameter $\ell$ on time advancement using terms up to the second PPN order $\varpropto aM$ and $M^{2} $ as the Bumblebee solution is valid only upto first order in $a$. It is shown that there is a critical radial distance $r_{c}$ above the Earth, where the Shapiro delay vanishes, and beyond $r_{c}$ the delay becomes negative, i.e., time advancement begins to set in, leading to the intriguing consequence that the measured LLR distance to Moon or any Satellite becomes \textit{less} than the zeroth order Euclidean distance. It is shown that the LSB correction arises from the conical geometry of the massless Bumblebee spacetime leading to upper bounds on the correction to the zeroth order Euclidean time interval as $\delta \tau _{\text{LSB}}^{\text{Eucl}}<0.8\times 10^{-4}$ (ns) and to time advancement as $\Delta \tau _{\text{LSB}}^{\text{adv}}<-4.5\times 10^{-13}$ (ns), both estimates based on the bound on $\ell $ corresponding to the Cassini spacecraft experiment. We shall also briefly touch upon the feasibility of direct experimental detection of the advancement effect.}


\maketitle

\section{Introduction}
\label{intro}
Gravitational time delay of signals, commonly known as Shapiro time delay \cite{Shapiro:1964}, is one of the well known experimental tests of Einstein's general relativity. The delay follows from the gravitational slowing down of signals as can be found from the integration of radial null geodesics in the Schwarzschild exterior metric modeling Sun's gravity. The subsequent experiment of Shapiro et al \cite{Shapiro:1971} measuring this delay involved radar signals that originated from the location of Earth (weak gravity of Sun), traveled to Mercury in superior conjunction grazing the rim of Sun (stronger gravity) and reflected back by Mercury to Earth along the same path. An Earth clock measured a proper time interval $\Delta \tau$ for the round trip, which yielded a positive excess time $\sim 240$ $\mu$sec over the round trip time along the corresponding Euclidean distance implying a time delay in excellent agreement with the prediction of general relativity. In contrast, an effect of gravitational\textit{\ time advancement }occurs in the reverse scenario, i.e., when the signal originates from a point in strong gravity, travels to a point in weaker gravity from where it is reflected back along the same path to the point of origin in the strong gravity. A clock located there then measures the proper time interval $\Delta \tau$ for the round trip that surprisingly yields a \textit{negative} excess time over the round trip time along the corresponding Euclidean distance, hence called time advancement. Rephrasing in terms of distance, a positive excess or delay implies that an extra distance (e.g., it is $\sim 72$ km in the Shapiro delay experiment \cite{Shapiro:1971}) is added to the Euclidean distance valid in the absence of gravitating mass. On the contrary, a negative excess or time advancement implies just the opposite - an extra distance is \textit{subtracted} from the Euclidean distance! Though at first sight this negative excess seems counterintuitive, this is yet another prediction of general relativity\footnote{%
For instance, the distance to Moon as measured by the Lunar Laser Ranging (LLR) technique could appear to us on Earth to be shorter by $22.09$ cm than the Euclidean distance of the zeroth order. It is a realizable physical prediction (see Sec.6).}. In what follows, it will be exemplifed that the time advancement effect could be a truly testable effect.

The phenomenon of time advancement was originally predicted in \cite{Bhadra:2010} for light traveling in the static exterior Schwarzschild spacetime as well as in a model independent centrally symmetric spacetime confirming the robustness of the effect. The possibility of its experimental observation in an Earth-Satellite configuration together with expected outcomes, measurable by current technology, was also suggested in \cite{Bhadra:2010}. Subsequently, time advancement effect under gravity's rainbow has been theoretically analyzed by Deng and Yi \cite{Deng:2017}. An interesting effect of the braneworld on the time advancement effect has been analyzed by Deng \cite{Deng:2018}. Most importantly, it was shown how the effect could be used to probe dark matter and dark energy distributions in the universe \cite{Ghosh:2015,Ghosh:2019}.

We wish to recall that, due to the universality of gravitation, clocks in a gravity field cannot measure the non-invariant coordinate time interval $\Delta t$ but can measure only the invariant proper time interval $\Delta \tau$. We further emphasize that time advancement (and for that matter, also the Shapiro delay) is measured by a clock located at a \textit{single} point in gravity from where the signal departs, and after reflection from a distant point, returns to the point of departure. A clock there measures the round-trip proper time interval between departure and arrival, which could show excess positive or negative time over the Euclidean time. Physically, the effects of advancement or Shapiro delay arise from the combined effect of gravitational slowing down of signals and the universality of gravitation manfested in the proper time as recorded by clocks located in stronger or in weaker gravity, as the case may be.

The Shapiro time delay was originally worked out up to first PPN order in Sun's mass using Schwarzschild exterior metric assuming it to model Sun's gravity \cite{Shapiro:1964}. Later, the delay has also been studied in the spherically symmetric black holes belonging to $f(T)$ gravity \cite{Farrugia:2016}. (See \cite{Carlip:2006} for some issues about the delay by Jupiter). However, since almost all astrophysical objects in the universe are spinning, it would be more appropriate to calculate the delay in a spinning exterior metric. To that end, the delay in the Kerr metric, modeling a spinning gravitating object of mass $M$, was first calculated by Krori \cite{Krori:1970} and Krori and Barua \cite{Krori:1986}, who found that the time delay of radar echo in the equatorial plane is independent of the spin $a$ to the first-order of approximation in mass ($\sim R_{s}=2M$). More recently, Wang and Lin \cite{Wang:2014} calculated the Shapiro time delay up to third PPN order in the Kerr metric obtaining an interesting result that the third-order mass term ($\sim R_{s}^{3}$) can \textit{exceed} the second order term ($\sim R_{s}a$), when the magnitude of impact factor $b$ is small.

With the above developments in mind, and inspired by the potential implications of the time advancement effect \cite{Bhadra:2010,Deng:2017,Ghosh:2015,Ghosh:2019}, we thought it worthwhile to explore the effect in the spinning solution of an exciting new theory of gravity, known as the Bumblebee gravity. The reason why it is exciting is as follows: The unification of classical gravity with other three fundamental forces of nature is one of the most important pursuits among the physics community and its successful realization would no doubt unravel new secrets of nature. However, physics near the Planck scale should necessarily take into account quantum principles in its essence. In this pursuit, models of quantum theories of gravity (QG) proposed so far have faced unsurmountable challenges since a direct verification of their predictions would require observing effects at the Planck scale energy ($\sim 10^{19}$ GeV) that is currently unavailable. An immediate possibility is to discard Lorentz invariance in the vicinity of the Planck scale where spacetime is discrete in nature. On the other hand, the standard model extension (SME), which is an effective field theory describing at low energies the general relativity and the standard model of particle physics, does include in its structure additional terms containing information about the Lorentz Symmetry Breaking (LSB) occurring near the Planck scale \cite{Kostelecky:2004}.

We wish to point out that LSB need not imply sacrifice of the physical content of well tested special relativity including the Michelson-Morley experiment \cite{Prokhovnik:1967,Prokhovnik:1985}. Lorentz symmetry, as commonly understood as reflecting the invariance of Minkowski spacetime, is actually a combination of \textit{real} physical effects of length contraction, time dilation, aberration etc and \textit{arbitrary convention} of clock synchronization. This important fact was first explained long ago by Tangherlini \cite{Tangherlini:1961,Tangherlini:2014}, further developed and analyzed later by Sj\"{o}din \cite{Sjodin:1979} and Ghosal et al \cite{Ghosal:1991}. On the other hand, physical effects cannot depend on arbitrary conventions. A synchronization is called standard or Einstein synchronization (characterized by a synchrony parameter $\xi =1$) in a given direction if it renders the one-way velocity of light in that direction equal to the one-way velocity in the opposite direction. In this case, the one-way velocity is equal to the round trip average velocity $c$ of light in vacuum. In this way, one defines an Einsteinian world with standard synchrony. Alternatively, one could synchronize clocks externally (characterized by $\xi =0$) that produces the same synchrony-independent special relativistic real effects \cite{Tangherlini:2014}. Note that, $\xi $ is conveniently normalized to the interval $0\leq \xi \leq 1$ and \textit{any} value of $\xi$ in this interval could be chosen but observable physical values nonetheless remain independent of $\xi $. For an explicit demonstration of this fact, see Sj\"{o}din \cite{Sjodin:1979}. However, see also \cite{Bhadra:2021}.

The idea of LSB as a novel way to probe for QG effects at low energy was first envisaged in the context of string theory \cite{Kostelecky:1989a,Kostelecky:1989b,Kostelecky:1989c,Colladay:1997}, non-commutative field theories \cite{Carroll:2001} or loop quantum gravity \cite{Gambini:1999,Ellis:2000}. Other important areas where LSB is applied include the neutrino sector \cite{Kostelecky:2004}, the extended standard model \cite{Dai:2017}, and the formation of atmospheric shower \cite{Rubtsov:2017}. The electromagnetic sector of the Standard Model Extension (SME) has been extensively studied in literature \cite{Bakke:2014,Kostelecky:1999,Yoder:2012,Lehnert:2003,Kharlanov:2007,Kostelecky:2001,Kostelecky:2002,Kostelecky:2006,Carroll:1990,Adam:2001,Chen:2000,Carone:2006,Klinkhamer:2011,Schreck:2012,Hohensee:2009,Altschul:2005}; the electroweak sector in \cite{Colladay:2009,Mouchrek:2017}, some aspects of the strong sector \cite{Abazov:2012}, the hadronic sector \cite{Berger:2016}. Furthermore, some effects of LSB in the gravitational sector have also been studied in \cite{Bluhm:2005,Maluf:2013,Maluf:2014,Bailey:2006}, specifically the case of the gravitational waves were analyzed in \cite{Kostelecky:2016a,Kostelecky:2016b}, Ricci dark energy in Bumblebee gravity model has been studied in \cite{Jesus:2019}. As a viable low energy probe of QG, Casana, Cavalcante, Poulis and Santos \cite{Casana:2017} recently proposed a Bumblebee vector field $B_{\mu}$ with a non-zero vacuum expectation value, $\left\langle B^{\mu}\right\rangle = b^{\mu }$, that is responsible for LSB. In their model, the Bumblebee vector field is coupled to a Riemann spacetime, which yielded a static spherically symmetric Schwarzschild-like black hole. They then studied the effects of Bumblebee parameter $\ell$ on some weak field solar system tests including Shapiro time delay and based on the experimental error incurred in measurements, obtained upper bounds on $\ell$ specific to experiments. In the backdrop of these develeopments, it would be desirable to study the effect of $\ell$ on time advancement in a spinning spacetime.

The present work will first extend the original time advancement effect \cite{Bhadra:2010} to spinning black holes ($a\neq 0$) up to third PPN order and then apply it to the Kerr-like black hole recently derived by Ding, Lui, Casana and Cavalcante \cite{Ding:2020} in the Lorentz violating Bumblebee gravity. (A more general spinning solution has been developed by Jha and Rahaman \cite{Jha:2021}). Next, using two practical configurations, Earth-Moon and Earth-Satellite, we shall calculate the influence of the Bumblebee parameter $\ell$ on time advancement computing terms up to the second PPN order $\varpropto aM$ and $M^{2}$ as the solution in \cite{Ding:2020} is valid only upto first order in spin $a$. The third order terms involving $a^{2}$ are nonetheless analytically worked out here for the Kerr expression at the limit $\ell =0$ \textit{albeit} it numerically contributes values too low to be of any physical interest. These configurations will exemplify that the two effects, time advancement and Shapiro delay, differ not only in sign but also in magnitude allowing for a detection of the advancement effect. Finally, we shall briefly touch upon the possibilities of experimental detection of the advancement effect.

The paper is organized as follows. To explain the new spacetime in which we are going to enumerate the time advancement effect, we include a brief outline of Bumblebee gravity in Sec.2. In Sec.3, we shall analytically calculate the effect up to third PPN order involving mass and spin. Sec.4 enumerates the advancement effect on the signals traveling back and forth along the Earth-Moon and Earth-Satellite radial lines and the values are displayed in Tables 1 and 2 respectively. Sec.5 contains a brief discussion on the experimental feasibility for detecting the effect. Sec.6 concludes the paper. Throughout this paper, we work in units where $8\pi G=1$ and $c=1$, unless specifically restored.

\section{Outline of Bumblebee gravity}
\label{sec:2}
Bumblebee model of gravity, first considered in the context of string theory \cite{Kostelecky:1989b}, is the simplest extension of classical general relativity allowing spontaneous violation of Lorentz symmetry and diffeomorphism by a Bumblebee vector field $B^{\mu}$. Under a suitable potential, the vector field acquires a non-vanishing vacuum expectation value that induces the LSB, which in turn is accompanied by diffeomorphism violation. The potential is chosen so as to possess a minimum that ensures the breaking of $U(1)$ gauge symmetry. Bumblebee models involving nonzero torsion in more general context are investigated in Refs.\cite{Kostelecky:2004,Bluhm:2005}. Assuming zero torsion, the Bumblebee gravity considered here is a special class of theories with an action given by \cite{Maluf:2014,Casana:2017}:
\begin{eqnarray}
S_{\text{B}} &=&S_{\text{EH}}+S_{\text{LSB}}+S_{\text{m}}  \nonumber \\
S_{\text{EH}} &=&\left( \frac{1}{2}\right) \int \sqrt{-g}d^{4}x\left(R-2\Lambda \right)   \nonumber \\
S_{\text{LSB}} &=&\left( \frac{1}{2}\right) \int \sqrt{-g}d^{4}x\left[uR+s^{\mu \nu }R_{\mu \nu }+t^{\mu \nu \alpha \beta }R_{\mu \nu \alpha \beta}\right]
\end{eqnarray}%
where $S_{\text{EH}}$ and $S_{\text{m}}$ are respectively the Einstein-Hilbert and matter action and $u,$ $s^{\mu \nu }$ and $t^{\mu\nu\alpha\beta}$ are LSB tensors in the $S_{\text{LSB}}$. The tensors $s^{\mu\nu}$ and $t^{\mu\nu\alpha\beta}$ possess the same symmetries as the Ricci and Riemann tensors respectively and are considered to be traceless.

Following \cite{Kostelecky:2004,Bluhm:2005,Maluf:2014}, the Bumblebee model can be represented by action (1), when $t^{\mu\nu\alpha\beta}=0$ and
\begin{equation}
u = \frac{\xi }{4}B^{\mu }B_{\mu },\text{ }s^{\mu \nu} = \xi \left( B^{\mu}B^{\nu} - \frac{1}{4}g^{\mu\nu}B_{\sigma}B^{\sigma}\right)
\end{equation}%
With these definitions, the action $S_{\text{B}}$ controlling the dynamics of the Bumblebee field $B_{\mu}$ is written as \cite{Maluf:2014}
\begin{equation}
S_{\text{B}} = \int \sqrt{-g}d^{4}x\left[ -\frac{1}{4}B^{\mu \nu }B_{\mu \nu} + \frac{\xi }{2}B^{\mu }B^{\nu }R_{\mu \nu }-V\left( B_{\mu }B^{\mu }\pm b^{2}\right) \right]
\end{equation}%
where the bumblebee field strength is defined as%
\begin{equation}
B_{\mu \nu } = \partial _{\mu }B_{\nu } - \partial _{\nu }B_{\mu },
\end{equation}%
The functional form of the potential $V(B^{\mu })$ that induces LSB has the functional form \cite{Kostelecky:1989b,Maluf:2014} has to be introduced by hand, which is conventionally taken as
\begin{equation}
V\left( B^{\mu }\right) =\frac{\lambda }{2}\left( B_{\mu }B^{\mu }\pm b^{2}\right) ^{2},
\end{equation}%
where $\lambda $ and $b^{2}$ are positive real constants, the non-minimal coupling constant $\xi$ has mass dimension $\left\vert \xi \right\vert = - 2$ and the same for fields are $\left\vert B^{\mu }\right\vert =1,$ $\left\vert B^{\mu \nu }\right\vert =2$. Some qualitative features of this potential have been studied in Refs.\cite{Kostelecky:1989a,Kostelecky:1989b,Kostelecky:1989c,Bluhm:2005,Kostelecky:2005,Maluf:2014}. The vacuum expectation value follows when $V=0$
\begin{equation}
B_{\mu }B^{\mu }\pm b^{2} = 0,
\end{equation}%
and this is possible when%
\begin{equation}
\left\langle B^{\mu }\right\rangle =b^{\mu },
\end{equation}%
so that the length of the vector is ($\pm $ corresponds to timelike and spacelike vectors respectively):
\begin{equation}
b_{\mu }b^{\mu }=\pm b^{2}.
\end{equation}%
The non-null squared length $b^{2}\neq 0$ is responsible for LSB.

\section{Time delay to third PPN order}
\label{sec:3}
To recall the Shapiro time delay in the Schwarzschild case, recall the propagation equation of the radial null geodesic in the equatorial plane
\begin{equation}
\left( \frac{dr}{dt}\right) ^{2}=A(r)\left\{ 1-\frac{r_{0}^{2}}{r^{2}}\frac{A(r)}{A(r_{0})}\right\} ,
\end{equation}%
where $A(r)=1-\frac{2M}{r}$, $M$ is the asymptotic ADM mass, $r_{0}$ is the closest approach distance to the trajectory from the center of the gravitating source, say, the Sun$.$ This equation tells us that the coordinate speed of light $\frac{dr}{dt}$ would be less than what it would be in the absence of mass for which $A(r)=1$. Hence the terminology "slowing down of light in the gravity field". Integration of Eq.(9) between two radial points $r_{0}$ and $r$ yields the two way coordinate time delay
\begin{equation}
\Delta t=2t(r_{0}\rightarrow r)=2\int_{r_{0}}^{r}\left[ A(r)\left\{ 1-\frac{%
r_{0}^{2}}{r^{2}}\frac{A(r)}{A(r_{0})}\right\} \right] ^{-1/2}dr.
\end{equation}%
Clocks could be placed either in the strong field $r_{0}$ near the gravitating source or at any point $r$ in the weak field far away from $r_{0}$. However, coordinate time delay is unmeasurable in a gravity field, so actual measurement by the clocks should yield the delay only in terms of proper time for two-way motion as
\begin{equation}
\Delta \tau = 2\sqrt{A(r)}\Delta t.
\end{equation}%
When the clock is in the weak field, i.e., $r>>2M$ so that $A(r)\simeq 1$, one has $\Delta \tau =2\sqrt{1-\frac{2M}{r}}\Delta t\simeq 2\Delta t$ without involving much error. This $\Delta t$ is calculated from the integral (10) by PPN expanding the integrand such that $\Delta \tau =2\Delta t=2\Delta t^{\text{Eucl}}+2\delta t$. The excess time $2\delta t$ is always positive and is known in the literature as the Shapiro time delay. However, when the clock is in the strong field, where $A(r_{0})\neq 1$, the clock measures
\begin{equation}
\Delta \tau = 2\sqrt{A(r_{0})}\Delta t\simeq 2\Delta t^{\text{Eucl}} + 2\delta t.
\end{equation}%
In this case, the excess time $2\delta t$ need not always be positive. We want to calculate the delay up to third PPN order and show that vanishing of the formal expression for $2\delta t$ in Eq.(12) reveals a critical distance for reflection depending on which $2\delta t$ can be positive, zero or \textit{negative}, the last being called the time advancement \cite{Bhadra:2010}. This is the central idea to be adopted below.

We choose the Kerr-like Bumblebee black hole to be at the origin of coordinates with the metric in Boyer-Lindquist coordinates obtained by Ding et al \cite{Ding:2020}:
\begin{eqnarray}
d\tau_{\text{Ding et al}}^{2} &=& -\left(1 - \frac{2Mr}{\rho^{2}}\right) dt^{2} - \frac{4Mar\sqrt{1 + \ell}\sin ^{2}\theta}{\rho^{2}} dt d\phi  \nonumber \\
&& + \left(\frac{\rho^{2}}{\Delta}\right) dr^{2} + \rho^{2}d\theta^{2} + \frac{H\sin^{2}\theta}{\rho^{2}}d\phi ^{2}, \\
\rho^{2} &=& r^{2}+a^{2}\cos ^{2}\theta , \\
\Delta &=& \frac{r^{2}-2Mr}{1+\ell }+a^{2}, \\
H &=& \left[ r^{2}+(1+\ell )a^{2}\right] ^{2}-\Delta (1+\ell )^{2}a^{2}\sin^{2}\theta ,
\end{eqnarray}%
where $\ell = \xi b^{2}$ is the Bumblebee parameter or LSB parameter. The equations of motion for the Bumblebee field $b^{\mu}$ are
\begin{equation}
\nabla ^{\mu }b_{\mu \nu }-\frac{\ell }{b^{2}}b^{\mu }R_{\mu \nu }=0.
\end{equation}%
A caveat must be mentioned here. Very recently, Maluf and Muniz \cite{Maluf:2022} (see also \cite{Kanzi:2022,Ding:2021}) have shown an important result that the equation of motion does not yield zero on the right hand side but yields terms proportional to $a^{2}$:%
\begin{equation}
\nabla ^{\mu }b_{\mu \nu }-\frac{\ell }{b^{2}}b^{\mu }R_{\mu \nu }\varpropto
a^{2},
\end{equation}%
so that the metric and the equations of motion are valid only up to \textit{first order} in spin $a$. So in the metric (13), all $a^{2}$ terms should be actually set to zero keeping only first order term in $a$ that appears the coefficient of $dtd\phi$. Nonetheless, we intentionally retain $a^{2}$ terms in the metric for the exclusive purpose of obtaining the analytical expressions for the limiting Kerr metric at $\ell = 0$ for which there is no Bumblebee field and Eq.(17) is trivially satisfied for all orders of $a$.

We assume that the signal motion is taking place in the exterior equatorial plane of the Bumblebee black hole ($\theta =\pi /2$) so that the metric becomes
\begin{equation}
d\tau ^{2}=-A(r)dt^{2}+B(r)dr^{2}+C(r)d\phi ^{2}-D(r)dtd\phi
\end{equation}%
\begin{eqnarray}
A(r) &=& 1 - \frac{R_{\text{S}}}{r}, \\
B(r) &=& \left[ \left( \frac{1}{1 + \ell }\right) \left(1 - \frac{R_{\text{S}}}{r}\right) + \frac{a^{2}}{r^{2}}\right] ^{-1}, \\
C(r) &=& r^{2}\left[1 + \left(1 + \ell \right) \frac{a^{2}}{r^{2}}\left(1 + \frac{R_{\text{S}}}{r}\right) \right] , \\
D(r) &=& \frac{2\sqrt{1 + \ell}aR_{\text{S}}}{r},
\end{eqnarray}%
where $R_{\text{S}}=2M$ is the Schwarzschild radius. It reduces to the Casana et al \cite{Casana:2017} Schwarzschild-like metric when $a=0$ and some of its novel features were presented in \cite{Izmailov:2022} with a new interpretation that $\ell$ could be related to \textit{conical angle} subtended at the center.

From the null geodesic $d\tau^{2} = 0$ in the spinning metric (13), we have
\begin{equation}
\frac{dr}{dt} = \pm \left(\frac{A(r) - F(r)}{B(r)}\right) ^{1/2},
\end{equation}%
where the sign of $\pm$ is for getting away and approaching the black hole respectively, and $F(r)$ is defined as
\begin{equation}
F(r)=C(r)\left( \frac{d\phi }{dt}\right) ^{2}-D(r)\left( \frac{d\phi }{dt}%
\right) .
\end{equation}%
The change rate of the azimuthal angle $\frac{d\phi}{dt}$ can be related to two constants of the motion along the given geodesic: the energy and angular momentum of the photon in the Kerr field, which are defined as \cite{Farrugia:2016, Krori:1970}
\begin{equation}
E = A(r)\overset{\cdot }{t}+\frac{D(r)}{2}\overset{\cdot }{\phi },
\end{equation}
\begin{equation}
L = C(r)\overset{\cdot }{\phi }-\frac{D(r)}{2}\overset{\cdot }{t},
\end{equation}%
here $\theta = \pi /2$ is assumed, and the dot represents the derivative with respect to an affine path parameter $\tau$. Without loss of generality, we can normalize $\tau$ such that $\overset{\cdot }{t} = 1$ at infinity which sets $E = 1$. This is convenient because $\overset{\cdot }{r}$ is then $\pm 1$ asymptotically. In this case, the impact factor $b$, which is defined as $L/E$, just equals $L$. Therefore, we can re-write $\frac{d\phi}{dt}$ as follow
\begin{equation}
\frac{d\phi}{dt} = \frac{D(r) + 2A(r)L}{2C(r) - D(r)L}.
\end{equation}

Substituting $\frac{d\phi }{dt}$ of Eq.(28) into Eq.(25), we have
\begin{equation}
F(r) = C(r) \left[\frac{D(r) + 2A(r)L}{2C(r) - D(r)L}\right]^{2} - D(r)\left[\frac{D(r) + 2A(r)L}{2C(r) - D(r)L}\right] .
\end{equation}%
Integrating Eq.(24), we can obtain the time for the light traveling from $r_{0}$ (closest approach distance, assumed to be the radius of the gravitating object) away to $r$ (the reflector) as follows
\begin{equation}
t(r_{0}\rightarrow r)=\overset{r}{\underset{r_{0}}{\int }}\left[ \frac{%
1-F(r)/A(r)}{B(r)/A(r)}\right] ^{-1/2}dr.
\end{equation}%
Expanding the metric in terms of $\left( \frac{R_{S}}{r}\right) ^{p}\left(\frac{a}{r}\right) ^{q}$ with the restriction of $p+q\leq 3$, $A(r)$, $B(r)$, $C(r)$ and $D(r)$ do not change. $F(r)$ can be written as
\begin{eqnarray}
F(r) &=& \frac{L^{2}}{r^{2}} \left[1 - \frac{2R_{\text{S}}}{r} + \frac{R_{\text{S}}^{2}}{r^{2}} - \frac{a^{2}}{r^{2}}\left(1 + \ell \right) + \frac{2aLR_{\text{S}}\sqrt{1 + \ell }}{r^{3}} - \frac{4aLR_{\text{S}}\sqrt{1 + \ell }}{r^{4}}  \right. \nonumber \\
&&\left.+\frac{a^{2}R_{\text{S}}}{r^{3}}\left(1 + \ell \right)\right], \\
B(r) &=& \left(1 + \ell \right) \left[1 + \frac{R_{\text{S}}}{r} + \frac{R_{\text{S}}^{2}}{r^{2}} + \frac{R_{\text{S}}^{3}}{r^{3}} - \left(1 + \ell \right) \left\{\frac{a^{2}}{r^{2}} + \frac{2a^{2}R_{\text{S}}}{r^{3}}\right\}\right].
\end{eqnarray}%
The traveling time is dependent on $r_{0}$, which is usually unknown but can be related to the impact factor $b$ . At $r=r_{0}$, we have from the combination of Eqs. (26) and (27) that
\begin{equation}
b = \frac{L}{E} = \frac{-D(r_{0})+2C(r_{0})\frac{d\phi }{dt}\vert_{r=r_{0}}}{2A(r_{0})+D(r_{0})\frac{d\phi }{dt}\vert_{r=r_{0}}}.
\end{equation}%
On the other hand, $\frac{dr}{dt}\vert_{r=r_{0}}=0$ (since $r_{0}$ is the minimum or closest approach distance) and it follows from Eq.(24) that $A(r_{0})=F(r_{0})$. Therefore, $\frac{d\phi}{dt}\vert_{r=r_{0}}$ can be solved to obtain the following quadratic algebraic equation
\begin{equation}
A(r_{0}) = C(r_{0})\left(\frac{d\phi}{dt}\vert_{r=r_{0}}\right)^{2}-D(r_{0})\left( \frac{d\phi }{dt}\vert_{r=r_{0}}\right).
\end{equation}%
To the third order of $R_{\text{S}}/r_{0}$, we have $\left. {}\right\vert $%
\begin{eqnarray}
\left. \frac{d\phi}{dt}\right\vert _{r=r_{0}} &=& \frac{\sqrt{1+\ell }aR_{\text{S}}}{r_{0}^{3}}\pm \frac{1}{r_{0}}\left[ 1-\frac{R_{\text{S}}}{2r_{0}}-\frac{R_{\text{S}}^{2}}{8r_{0}^{2}}-\left( 1+\ell \right) \frac{a^{2}}{2r_{0}^{2}}-\left( 1+\ell \right) \frac{a^{2}R_{\text{S}}}{4r_{0}^{3}} \right. \nonumber \\
&&\left. - \frac{1}{16}\frac{R_{\text{S}}^{3}}{r_{0}^{3}}\right].
\end{eqnarray}%
The $\pm$ sign correspond to direct and retrograde orbits relative to the spin of the central object respectively.

Plugging Eq.(35) into Eq.(33), we can write $b$ in terms of $r_{0}$ as follow
\begin{eqnarray}
b=L&=&\pm r_{0}\left[ 1+\frac{R_{\text{S}}}{2r_{0}}+\frac{3R_{\text{S}}^{2}}{8r_{0}^{2}}\mp \frac{aR_{\text{S}}\sqrt{1+\ell }}{r_{0}^{2}}+\left( 1+\ell \right) \frac{a^{2}}{2r_{0}^{2}}+\frac{5R_{\text{S}}^{3}}{16r_{0}^{3}} \right. \nonumber \\
&&\left. \mp \frac{aR_{\text{S}}^{2}\sqrt{1+\ell }}{r_{0}^{3}}+\left( 1+\ell \right) \frac{3a^{2}R_{\text{S}}}{4r_{0}^{3}}\right]
\end{eqnarray}%
Substituting Eq.(36) into Eq.(31), we have
\begin{eqnarray}
F(r,r_{0}) &=&\frac{r_{0}^{2}}{r^{2}}\left[ 1+\frac{r-2r_{0}}{rr_{0}}R_{%
\text{S}}+\frac{(r-r_{0})^{2}}{r^{2}r_{0}^{2}}R_{\text{S}}^{2}\mp \frac{2%
\sqrt{1+\ell }(r^{3}-r_{0}^{3})}{r^{3}r_{0}^{2}}aR_{\text{S}} \right. \nonumber \\
&& +\frac{\left(1+\ell \right) \left( r^{2}-r_{0}^{2}\right) }{r^{2}r_{0}^{2}}a^{2}+\frac{%
(r-r_{0})^{2}}{r^{2}r_{0}^{3}}R_{\text{S}}^{3}  \nonumber \\
&&\mp \frac{\sqrt{1+\ell }(3r-4r_{0})(r^{3}-r_{0}^{3})}{r^{4}r_{0}^{3}%
}aR_{\text{S}}^{2}  \nonumber \\
&&\left. +\frac{\left( 1+\ell \right) (r-r_{0})(2r^{2}-r_{0}^{2})}{%
r^{3}r_{0}^{3}}a^{2}R_{\text{S}}\right].
\end{eqnarray}%
Inserting Eqs.(20), (32) and (37) into Eq.(30), we have
\begin{eqnarray}
t(r_{0}\rightarrow r) &=&\sqrt{1+\ell }\int\limits_{r_{0}}^{r}\left( \frac{%
r^{2}}{r^{2}-r_{0}^{2}}\right) ^{\frac{1}{2}}\left[ 1+\frac{2r+3r_{0}}{%
2r(r+r_{0})}R_{\text{S}}+\frac{3(4r^{2}+8rr_{0}+5r_{0}^{2})}{%
8r^{2}(r+r_{0})^{2}}R_{\text{S}}^{2}\right.  \notag \\
&& \mp \frac{\sqrt{1+\ell}(r^{2}+rr_{0}+r_{0}^{2})}{r^{3}(r+r_{0})}aR_{\text{S}} + \frac{8r^{4}+10r^{3}r_{0}+92r^{2}r_{0}^{2}+90rr_{0}^{3}+35r_{0}^{4}}{16r^{3}r_{0}(r+r_{0})^{3}}R_{\text{S}}^{3} \notag \\
&&\left. \mp \frac{3\sqrt{1+\ell }%
(r^{2}+rr_{0}+r_{0}^{2})^{2}}{2r^{4}r_{0}(r+r_{0})^{2}}aR_{\text{S}}^{2}+%
\frac{1+\ell }{rr_{0}(r+r_{0})}a^{2}R_{\text{S}}\right] dr.
\end{eqnarray}%
Calculating the integration gives final result:
\begin{eqnarray}
t(r_{0} &\rightarrow &r) = \Delta t = \sqrt{1 + \ell}\left\{\sqrt{r^{2} - r_{0}^{2}} + \left[ \left( \frac{1}{2}\right) \sqrt{\frac{r - r_{0}}{r + r_{0}}} + \ln{\left( \frac{r + \sqrt{r^{2} - r_{0}^{2}}}{r_{0}}\right) }\right] R_{\text{S}} \right.  \nonumber \\
&& + \left[\frac{15}{8r_{0}}\arccos {\left( \frac{r_{0}}{r}\right) } - \frac{(4r+5r_{0})\sqrt{r-r_{0}}}{8r_{0}(r+r_{0})^{\frac{3}{2}}}\right] R_{\text{S}}^{2}  \mp \frac{\sqrt{1+\ell }(2r+r_{0})\sqrt{r-r_{0}}}{rr_{0}\sqrt{r+r_{0}}}aR_{\text{S}} \nonumber \\
&&+\left[ \frac{\sqrt{r^{2}-r_{0}^{2}}(60r^{3}+157r^{2}r_{0}+133rr_{0}^{2}+35r_{0}^{3})}{16rr_{0}^{2}(r+r_{0})^{3}} + \frac{15}{16r_{0}^{2}}\arccos {\left( \frac{r_{0}}{r}\right) }\right] R_{\text{S}}^{3}  \nonumber \\
&& \pm \sqrt{1+\ell }\left[ \frac{\sqrt{r^{2}-r_{0}^{2}}(8r^{3}+7r^{2}r_{0}-6rr_{0}^{2}-3r_{0}^{3})}{4r^{2}r_{0}^{2}(r+r_{0})^{2}} + \frac{15}{4r_{0}^{2}}\arccos {\left( \frac{r_{0}}{r}\right) }\right] aR_{\text{S}}^{2} \nonumber \\
&&\left.  +\frac{(1+\ell )\sqrt{r-r_{0}}}{r_{0}^{2}\sqrt{r+r_{0}}}a^{2}R_{\text{S}} \right\}.
\end{eqnarray}

The above is the Shapiro delay up to third PPN order\footnote{%
When $\ell = 0$, the cubic Kerr terms in $T(r)$ here differ from those of Wang and Lin \cite{Wang:2014}. However, their conclusion that the third-order mass effect is larger than the rotation effect for certain ranges of impact parameter is correct.}. We shall compute only up to second order mass term $R_{\text{S}}^{2}$ and only the first order spin term (i.e., $aR_{\text{S}}$) since the third order terms in $R_{\text{S}}^{3}$ is too miniscule to be worth considering\footnote{%
When $\ell \neq 0$, the third order term involving $a^{2}$, i.e., $a^{2}R_{\text{S}}$ is \textit{not }to be considered at all since the Ding et al \cite{Ding:2020} metric (13-16) is valid only up to first order in spin $a$ \cite{Maluf:2022}. However, the expressions for $t(r_{0}\rightarrow r)$ are valid for all order in $a$ \textit{only} when $\ell = 0$ (Kerr metric of general relativity).}.

\section{Practical applications}
\label{sec:4}
As a practical application, we assume an idealized situation that the signal originates in the central gravity field of the Earth ($B$), and after being reflected by the Moon ($A$), returns along the same path to the point of origin on Earth ($B$). As in Fig.1, assume that $r=r_{A}$, the distance between the Earth and the Moon; $r_{0}=r_{B}$, the radius of the Earth; $M = M_{\oplus }$, the mass of the Earth; $R_{\text{S}} = 2M_{\oplus }$, $a = a_{\oplus }$; the spin of the Earth and $\Delta t$ is the two-way coordinate time difference between emission and arrival. The total difference in proper time between transmission and reception of the signal to be measured by a clock at $B$ (Earth), following Eq.(11), is
\begin{equation}
\Delta \tau _{BAB}^{\text{total}}=2\left( \sqrt{1-\frac{2M_{\oplus }}{r_{B}}}%
\right) \Delta t\simeq 2\left( 1-\frac{M_{\oplus }}{r_{B}}-\frac{M_{\oplus
}^{2}}{2r_{B}^{2}}\right) \Delta t
\end{equation}%
where the one-way coordinate time difference up to second PPN order is, from
Eq.(39),
\begin{eqnarray}
\Delta t &=& \sqrt{1 + \ell }\sqrt{r_{A}^{2} - r_{B}^{2}} + 2\sqrt{1 + \ell}M_{\oplus}\left[\ln\left( \frac{r_{A} +\sqrt{r_{A}^{2} - r_{B}^{2}}}{r_{B}}\right) + \left(\frac{1}{2}\right) \sqrt{\frac{r_{A} - r_{B}}{r_{A} + r_{B}}}\right]   \nonumber \\
&&+\frac{\sqrt{1+\ell }M_{\oplus }^{2}}{2r_{B}}\left[ 15\arccos\left(\frac{r_{B}}{r_{A}}\right) -\sqrt{\frac{r_{A}-r_{B}}{r_{A}+r_{B}}}\left(\frac{4r_{A}+5r_{B}}{r_{A}+r_{B}}\right) \right]   \nonumber \\
&&\mp (1 + \ell )\left( \frac{2M_{\oplus }a_{\oplus }}{r_{B}}\right) \left[\sqrt{\frac{r_{A}-r_{B}}{r_{A}+r_{B}}}+\frac{\sqrt{r_{A}^{2}-r_{B}^{2}}}{r_{A}}\right]
\end{eqnarray}

\begin{figure}[!ht]
  \centerline{\includegraphics[scale=1.5]{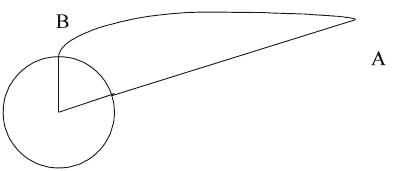}}
  \caption{The configuration showing the gravitating source and the distances of the observer at $B$ and the reflector at $A$.}
  \label{fig1}
\end{figure}

The Shapiro effect measured by an observer on Earth after the two-way motion of signal sent from Earth to Moon and reflected back to Earth (two-way) can be found using Eqs.(40) and (41), and written out separating the mass and spin effects up to second order as
\begin{equation}
\Delta \tau _{BAB}^{\text{total}}=\Delta \tau _{0}+\Delta \tau _{(M_{\oplus
})}+\Delta \tau _{(M_{\oplus }^{2})}+\Delta \tau _{(M_{\oplus }a_{\oplus })},
\end{equation}%
where the components are
\begin{eqnarray}
\Delta \tau_{0} &=& 2\sqrt{1 + \ell }\sqrt{r_{A}^{2} - r_{B}^{2}}, \\
\Delta \tau _{(M_{\oplus })} &=& 4\sqrt{1 + \ell }M_{\oplus }\left[\ln \left(\frac{r_{A} + \sqrt{r_{A}^{2} - r_{B}^{2}}}{r_{B}}\right) +\left( \frac{1}{2}\right) \sqrt{\frac{r_{A}-r_{B}}{r_{A}+r_{B}}}\right]  \nonumber \\
&& - 2\sqrt{1 + \ell }M_{\oplus }\frac{\sqrt{r_{A}^{2} - r_{B}^{2}}}{r_{B}}, \\
\Delta \tau_{(M_{\oplus }^{2})} &=& \frac{\sqrt{1+\ell }M_{\oplus }^{2}}{r_{B}}\left[15\arccos\left( \frac{r_{B}}{r_{A}}\right) - \sqrt{\frac{r_{A}-r_{B}}{r_{A}+r_{B}}}\left( \frac{4r_{A}+5r_{B}}{r_{A}+r_{B}}\right)\right]  \nonumber \\
&& -2\sqrt{1+\ell }M_{\oplus }^{2}\frac{\sqrt{r_{A}^{2}-r_{B}^{2}}}{2r_{B}^{2}}  \nonumber \\
&&-\frac{4\sqrt{1+\ell }M_{\oplus }^{2}}{r_{B}}\left[ \ln \left( \frac{r_{A} + \sqrt{r_{A}^{2}-r_{B}^{2}}}{r_{B}}\right) +\left( \frac{1}{2}\right) \sqrt{\frac{r_{A}-r_{B}}{r_{A}+r_{B}}}\right],  \\
\Delta \tau _{(M_{\oplus }a_{\oplus })} &=&\mp (1+\ell )\left( \frac{4M_{\oplus }a_{\oplus }}{r_{B}}\right) \left[ \sqrt{\frac{r_{A}-r_{B}}{r_{A}+r_{B}}}+\frac{\sqrt{r_{A}^{2}-r_{B}^{2}}}{r_{A}}\right] .
\end{eqnarray}%
It is the last term $\left( -2\sqrt{1+\ell }M_{\oplus }\frac{\sqrt{r_{A}^{2}-r_{B}^{2}}}{r_{B}}\right) $ in Eq.(44) that is negative and large dominating over all the other terms in that expression as well as other higher order PPN terms such that $\Delta \tau _{BAB}^{\text{total}} - \Delta \tau _{0}=\Delta \tau _{(M_{\oplus })}+\Delta \tau _{(M_{\oplus}^{2})}+\Delta \tau _{(M_{\oplus }a_{\oplus })}<0$, an effect that we denote by \textit{time advancement}. Intriguingly, there now appears a critical radius $r=r_{c}$ between the Earth and Moon (Fig.2) where $\Delta \tau _{BAB}^{\text{total}}-\Delta \tau _{0}=0$. At $r<r_{c}$, $\Delta \tau _{BAB}^{\text{total}}-\Delta \tau _{0}>0$ (Shapiro delay); at $r>r_{c}$, $\Delta \tau _{BAB}^{\text{total}}-\Delta \tau _{0}<0$ (time advancement). The higher order terms $\Delta \tau _{(M_{\oplus }^{2})}$ and $\Delta \tau _{(M_{\oplus }a_{\oplus })}$ in Eqs.(45) and (46) are also negative but far too miniscule as Table 1 shows though the spin effect $\Delta \tau _{(M_{\oplus }a_{\oplus })}$ can be hopefully measurable in the near future at the femtosecond level at $\ell =0$.

\begin{figure}[!ht]
  \centerline{\includegraphics[scale=1.05]{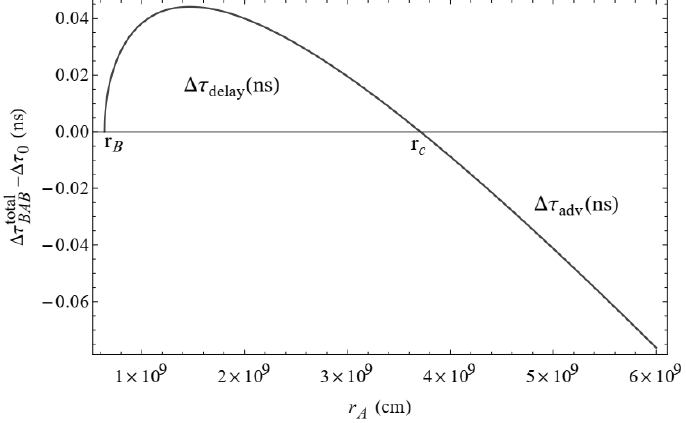}}
  \caption{The two-way Shapiro time delay [Eq.(42)]: $\Delta \tau _{\scriptsize{\textmd{delay}}%
}=\Delta \tau _{BAB}^{\scriptsize{\textmd{total}}}-\Delta \tau _{0}=\Delta \tau
_{(M_{\oplus })}+\Delta \tau _{(M_{\oplus }^{2})}+\Delta \tau _{(M_{\oplus
}a_{\oplus })}$ measured by an Earth clock at $r_{B}$. The signal travels
from a point on Earth's surface $r_{B}$ and is reflected back by a
reflector, stationed at an arbitrary distance $r_{A}$, to the same point on
Earth, where a clock measures the delay in the following manner: (i) If the
reflector is at a distance $r_{A}<$ $r_{c}$, the Earth clock will measure
the usual positive Shapiro time delay, $\Delta \tau _{\scriptsize{\textmd{delay}}}>0$. (ii)
If the reflector is at a distance $r_{A}=r_{c}$ (critical point), then the
the Earth clock will observe neither delay nor advancement, $\Delta \tau _{%
\scriptsize{\textmd{delay}}}=0$. (iii) If the reflector is at a distance $r_{A}>$ $r_{c}$,
the Earth clock will measure a negative Shapiro delay $\Delta \tau _{\scriptsize{\textmd{%
delay}}}<0$ or gravitational time advancement.}
  \label{fig2}
\end{figure}

Assume, on the other hand, that the signals were sent from the Moon ($r_{A}$) to a reflector on Earth ($r_{B}$) that is sending the signal back to Moon ($r_{A}$). Then the Moon clock will measure the usual Shapiro delay to be
\begin{eqnarray}
\Delta \tau _{ABA}^{\text{total}} &=&2\left( \sqrt{1-\frac{2M_{\oplus }}{%
r_{A}}}\right) \Delta t\simeq 2\left( 1-\frac{M_{\oplus }}{r_{A}}-\frac{%
M_{\oplus }^{2}}{2r_{A}^{2}}\right) \Delta t  \nonumber \\
&=&\Delta \tau _{0}+\Delta \tau _{(M_{\oplus })}+\Delta \tau _{(M_{\oplus
}^{2})}+\Delta \tau _{(M_{\oplus }a_{\oplus })},
\end{eqnarray}%
which yields the components
\begin{eqnarray}
\Delta \tau _{0} &=&2\sqrt{1+\ell }\sqrt{r_{A}^{2}-r_{B}^{2}}, \\
\Delta \tau _{(M_{\oplus })} &=&4\sqrt{1+\ell }M_{\oplus }\left[ \ln \left(\frac{r_{A}+\sqrt{r_{A}^{2}-r_{B}^{2}}}{r_{B}}\right) +\left( \frac{1}{2}\right) \sqrt{\frac{r_{A}-r_{B}}{r_{A}+r_{B}}}\right] \nonumber \\
&&-2\sqrt{1+\ell }M_{\oplus }\frac{\sqrt{r_{A}^{2}-r_{B}^{2}}}{r_{A}}, \\
\Delta \tau _{(M_{\oplus }^{2})} &=&\frac{\sqrt{1+\ell }M_{\oplus }^{2}}{r_{B}}\left[ 15\arccos \left( \frac{r_{B}}{r_{A}}\right) -\sqrt{\frac{r_{A}-r_{B}}{r_{A}+r_{B}}}\left( \frac{4r_{A}+5r_{B}}{r_{A}+r_{B}}\right)\right]  \nonumber \\
&&-2\sqrt{1+\ell }M_{\oplus }^{2}\frac{\sqrt{r_{A}^{2}-r_{B}^{2}}}{2r_{A}^{2}}  \nonumber \\
&&-\frac{4\sqrt{1+\ell }M_{\oplus }^{2}}{r_{B}}\left[ \ln \left( \frac{r_{A} + \sqrt{r_{A}^{2}-r_{B}^{2}}}{r_{B}}\right) +\left( \frac{1}{2}\right) \sqrt{\frac{r_{A}-r_{B}}{r_{A}+r_{B}}}\right],  \\
\Delta \tau _{(M_{\oplus }a_{\oplus })} &=&\mp (1+\ell )\left( \frac{4M_{\oplus }a_{\oplus }}{r_{B}}\right) \left[ \sqrt{\frac{r_{A}-r_{B}}{r_{A}+r_{B}}}+\frac{\sqrt{r_{A}^{2}-r_{B}^{2}}}{r_{A}}\right] .
\end{eqnarray}

\begin{figure}[!ht]
  \centerline{\includegraphics[scale=1.05]{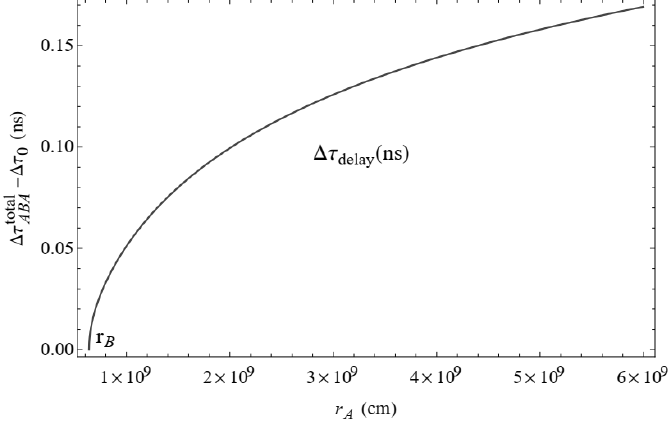}}
  \caption{The two-way Shapiro time delay, $\Delta \tau _{\scriptsize{\textmd{delay}}} = \Delta \tau _{ABA}^{\scriptsize{\textmd{total}}} - \Delta \tau _{0} = \Delta \tau _{(M_{\oplus})}+\Delta \tau _{(M_{\oplus }^{2})}+\Delta \tau_{(M_{\oplus }a_{\oplus})}>0$, measured by a clock at the location $r_{A}$, when the signal emerges from $r_{A}$, and after being reflected at $r_{B}$, returns to $r_{A}$. This is the usual positive Shapiro delay phenomenon.}
  \label{fig3}
\end{figure}

Here the last term in Eq.(49), viz., $-2\sqrt{1+\ell }M_{\oplus }\frac{\sqrt{r_{A}^{2}-r_{B}^{2}}}{r_{A}}$, though negative, is too small to offset the preceding positive term or other positive terms in Eqs.(50) and (51). As shown in Fig.3, the $\Delta \tau _{ABA}^{\text{total}}-\Delta \tau _{0}=\Delta \tau _{(M_{\oplus })}+\Delta \tau _{(M_{\oplus }^{2})}+\Delta \tau _{(M_{\oplus }a_{\oplus })}>0$ in exact accordance with the usual Shapiro delay as expected in the redefined Moon-Earth configuration. Note from Eqs.(43) and (48) that the Euclidean distance $\sqrt{r_{A}^{2}-r_{B}^{2}}$ is modified by a factor $\sqrt{1+\ell }$ but this is expected as the spacetime does not become flat or planar in the equatorial section when $M=0$ and $a=0.$ In fact, the $(r,\phi )$ metric becomes a metric on a \textit{conical} surface since $B=1+\ell $ and $C(r)=r^{2}$. As a consequence, the two-way Euclidean time interval is distorted by the factor $\sqrt{1+\ell }$ yielding an LSB correction $\delta \tau _{\text{LSB}}^{\text{Eucl}}$ to the undistorted Euclidean distance (see Appendix):
\begin{eqnarray}
\Delta \tau _{0} &=&2\sqrt{1+\ell }\left( \sqrt{r_{A}^{2}-r_{B}^{2}}\right) \simeq 2\left( 1+\frac{\ell }{2}\right) \left( \sqrt{r_{A}^{2}-r_{B}^{2}}\right)  \nonumber \\
&=& 2\sqrt{r_{A}^{2}-r_{B}^{2}}+\ell \sqrt{r_{A}^{2}-r_{B}^{2}} \\
&=& \delta \tau _{0}+\delta \tau _{\text{LSB}}^{\text{Eucl}}.
\end{eqnarray}%
Restoring $c$, the Lorentz violating (LSB) correction $\delta \tau _{\text{LSB}}$ to the Earth-Moon Euclidean distance, viz., $c\delta \tau _{0}=\sqrt{r_{A}^{2}-r_{B}^{2}}$, thus becomes
\begin{equation}
\delta \tau _{\text{LSB}}^{\text{Eucl}}=\frac{\ell }{c}\sqrt{r_{A}^{2}-r_{B}^{2}}=\frac{\ell }{2}\delta \tau _{0}=\left( \frac{\ell }{2}\right) \times 2.56\text{ sec},
\end{equation}%
the middle expression $\frac{\ell }{2}\delta \tau _{0}$ is precisely what Casana et al \cite{Casana:2017} obtained as the LSB contribution. With the most accurate (with maximum error $0.0012\%$ of unity) value from the Cassini spacecraft experiment \cite{Bertotti:2003}, Casana et al \cite{Casana:2017} deduced $\ell _{\text{Cassini}}<6.2\times 10^{-13}$, and using this we find%
\begin{equation}
\delta \tau _{\text{LSB}}^{\text{Eucl}}<0.8\times 10^{-4}\text{ ns.}
\end{equation}%
This upper limit is magnitudewise far less than the first order mass contribution to time advancement $\Delta \tau _{(M_{\oplus })}$, viz., $-1.460$ ns but should still be measurable once we deduct the pure Euclidean value $\delta \tau _{0}=2.56$ sec in Eq.(53). The effect of $\ell$ on the time advancement influencing the leading order term $\Delta \tau_{(M_{\oplus })}$ in Eq.(44) is much smaller than unity. A rough estimate using the Cassini bound suggests a bound on the LSB correction to this term as
\begin{eqnarray}
\Delta \tau _{(M_{\oplus })\text{LSB}}^{\text{adv}} &=&\left( \frac{\ell }{2}\right) \left( 4M_{\oplus }\left[ \ln \left( \frac{r_{A}+\sqrt{%
r_{A}^{2}-r_{B}^{2}}}{r_{B}}\right) +\left( \frac{1}{2}\right) \sqrt{\frac{r_{A}-r_{B}}{r_{A}+r_{B}}}\right] \right.  \nonumber \\
&&\left.-2M_{\oplus }\frac{\sqrt{r_{A}^{2}-r_{B}^{2}}}{r_{B}}\right) \\
&<&-4.526\times 10^{-13}\text{ ns,}
\end{eqnarray}%
the measurement of which seems technologically out of question today but it is still noteworthy to mention that today a time interval can be measured with an accuracy better than a femtosecond \cite{OBrian:2004}.

\begin{table}[!ht]
\centering
\begin{tabular}{|c|c|c|c|}
\hline
$\ell $ & $\Delta \tau _{(M_{\oplus })}$ (ns) & $\Delta \tau _{(M_{\oplus
}^{2})}$ (ns) & $\Delta \tau _{(M_{\oplus }a_{\oplus })}$ (ns) \\ \hline
$0$ & $-1.460$ & $-6.291\times 10^{-10}$ & $-5.766\times 10^{-6}$ \\
$0.01$ & $-1.467$ & $-6.322\times 10^{-10}$ & $-5.824\times 10^{-6}$ \\
$0.05$ & $-1.496$ & $-6.446\times 10^{-10}$ & $-6.054\times 10^{-6}$ \\
$0.1$ & $-1.531$ & $-6.598\times 10^{-10}$ & $-6.342\times 10^{-6}$ \\
$0.2$ & $-1.599$ & $-6.891\times 10^{-10}$ & $-6.919\times 10^{-6}$ \\
$0.3$ & $-1.664$ & $-7.172\times 10^{-10}$ & $-7.496\times 10^{-6}$ \\
$0.4$ & $-1.727$ & $-7.443\times 10^{-10}$ & $-8.072\times 10^{-6}$ \\
$0.5$ & $-1.788$ & $-7.704\times 10^{-10}$ & $-8.649\times 10^{-6}$ \\
$1$ & $-2.064$ & $-8.896\times 10^{-10}$ & $-1.153\times 10^{-5}$ \\
$2$ & $-2.528$ & $-1.090\times 10^{-9}$ & $-1.729\times 10^{-5}$ \\
\hline
\end{tabular}
\caption{We show the comparison of different parts of two-way time advancement $\Delta \tau _{BAB}^{\scriptsize{\textmd{total}}}-\Delta \tau _{0}$ [Eq.(42)] undergone by the signal sent from the Earth to Moon and reflected back to Earth for arbitrary values of $\ell $. For numerical estimates we have used typical values: distance between Earth and Moon $r_{A}=3.844\times 10^{10}$ cm; radius of the Earth $r_{B}=6.371\times 10^{8}$ cm; mass of the Earth $M_{\oplus }=0.44$ cm and spin of the Earth $a_{\oplus }=3.27\times 10^{2}$ cm. The values are shown in nano-second (ns).}
\end{table}

\begin{table}[!ht]
\centering
\begin{tabular}{|c|c|c|c|}
\hline
$\ell $ & $\Delta \tau _{(M_{\oplus })}$ (ps) & $\Delta \tau _{(M_{\oplus
}^{2})}$ (ps) & $\Delta \tau _{(M_{\oplus }a_{\oplus })}$ (ps) \\ \hline
$0$ & $-41.233$ & $-2.546\times 10^{-8}$ & $-5.440\times 10^{-3}$ \\
$0.01$ & $-41.438$ & $-2.559\times 10^{-8}$ & $-5.495\times 10^{-3}$ \\
$0.05$ & $-42.251$ & $-2.609\times 10^{-8}$ & $-5.712\times 10^{-3}$ \\
$0.1$ & $-43.245$ & $-2.670\times 10^{-8}$ & $-5.985\times 10^{-3}$ \\
$0.2$ & $-45.168$ & $-2.789\times 10^{-8}$ & $-6.529\times 10^{-3}$ \\
$0.3$ & $-47.013$ & $-2.903\times 10^{-8}$ & $-7.073\times 10^{-3}$ \\
$0.4$ & $-48.787$ & $-3.012\times 10^{-8}$ & $-7.617\times 10^{-3}$ \\
$0.5$ & $-50.500$ & $-3.118\times 10^{-8}$ & $-8.161\times 10^{-3}$ \\
$1$ & $-58.312$ & $-3.601\times 10^{-8}$ & $-1.088\times 10^{-2}$ \\
$2$ & $-71.418$ & $-4.410\times 10^{-8}$ & $-1.632\times 10^{-2}$ \\
\hline
\end{tabular}
\caption{We show the comparison of different parts of two-way time advancement $\Delta \tau _{BAB}^{\scriptsize{\textmd{total}}} -\Delta \tau _{0}$ [Eq.(42)] undergone by the signal sent from the Earth to a Satellite and reflected back to Earth for arbitrary values of $\ell $. For numerical estimates we have used typical values: distance between the Earth and the Satellite $r_{A}=5\times 10^{9}$ cm; radius of the Earth $r_{B}=6.371\times 10^{8}$ cm; mass of the Earth $M_{\oplus }=0.44$ cm and spin of the Earth $a_{\oplus }=3.27\times 10^{2}$ cm. The values are shown in pico-second (ps).}
\end{table}

\section{Feasibility of experimental detection of the effect}
\label{sec:5}
A key question at this stage is the feasibility of experimental detection of gravitational time advancement in the Earth-Moon system. Since 1969 the lunar laser ranging (LLR) experiments are measuring the round-trip travel time of laser pulses between the Earth surface and the reflectors (retroreflector arrays) placed on the lunar surface by the Apollo astronauts, and by an unmanned Soviet rover. Besides numerous new information about the dynamics and structure of the Moon, the LLR data has been successfully employed to test a number of basic gravitational aspects, which include the time variation of gravitational constant, testing of equivalence principle, geodetic precession and the PPN parameters in the barycentric system, where the motion of Earth is simulated as a motion of an
$N-$body potential hump \cite{OBrian:2004,Battat:2009, Muller:2019,Mazarico:2020,Viswanathan:2018}. However, for the purpose of illustrating our idea of assessing the LSB effect in the time advancement phenomenon, we consider the geocentric system in which the Moon is moving as a test object around the Earth. The numerical values computed in the tables refer to this configuration only.

Initially the precision of LLR was a few hundred centimeters, which has been improved to a few millimeters in recent times \cite{Battat:2009, Muller:2019}. The Apache point lunar laser-ranging operation (APOLLO) measures the round trip time interval to few-picosecond precision, which means that the present generation LLR observations can cleanly detect the effect of time advancement provided one knows the Euclidean distance between the Earth and the Moon from any other independent observations with high precision, which is currently not available. One can also test the time advancement effect if Earth laser ranging (ELR) experiments can be operated from the Moon surface; in that case, the difference in time intervals of the LLR and ELR will give the time advancement effect. Even if one takes the LLR instruments on the Moon surface by spacecraft etc, the ELR experiment from there (the moon surface) is still difficult to operate, particularly for the environmental conditions of the Moon.

Very recently, an LLR experiment has been conducted from the ground station in Grasse, France, exploiting a spacecraft at lunar distance, the Lunar Reconnaissance Orbiter (LRO) \cite{Mazarico:2020}. To operate an ELR experiment from a spacecraft will be a challenging task, mainly because of (relatively) small area available on the spacecraft for hosting the LLR equipment. Since the Moon/spacecraft has no atmosphere the beam divergence will be much less but the return beam after bouncing back from a reflector on the Earth surface will be diverged due to Earth's atmosphere which may be minimized by placing the reflectors at high altitude (on high mountain, for example). With today's more sophisticated laser ranging systems, the precision of the Earth-Moon distance measurement is now $\sim 0.01$ mm \cite{Viswanathan:2018}, $25$ times better than the first calculations in 1969 measured using the 3-meter telescope at California's Lick Observatory. The proposed mission Beyond Einstein Advanced Coherent Optical Network (BEACON) will employ four small spacecrafts equipped with laser transceivers. From a circular Earth orbit it will measure the distances between the spacecraft to very high accuracy ($\sim 0.1$ nm). So it appears that we have the technology to conduct ELR from a spacecraft.

\section{Conclusions}
\label{concl}
An important caveat was pointed out in \cite{Maluf:2022} that the Ding et al \cite{Ding:2020} metric is not an exact solution of the equations of motion except only in the slow spin limit, i.e., to first order in spin $a$. The Jha and Rahaman \cite{Jha:2021} solution that generalizes the Ding et al \cite{Ding:2020} solution also suffers from the same malady pointed out in \cite{Maluf:2022} and admitted by the authors of \cite{Jha:2021}. This being the case, it is necessary to provide some reasonable justifications as to why the Ding et al metric was considered at all. There are, in principle, three reasons as to why the first order spin effect could still capture much of its contribution. The first is that the spin effect decays faster than the mass effect in the weak field limit, so probably not much of spin contribution will be left unaccounted for if terms in $a^{2}$ and higher are neglected. The second reason is that the Ding et al metric in the first order in $a$ can be treated as having the same status as the well known Lense-Thirring metric, a counterpart of the Kerr metric in the same approximation that produces, for example, the frame-dragging effect. These facts encouraged us to extend the analysis of the time advancement effect, initially developed for the Schwarzschild metric \cite{Bhadra:2010} and recently applied to Bumblebee static metric \cite{Izmailov:2022}, to a spinning metric to first order in $a$, as was tabulated above. The work also automatically reveals the time advancement effect in the spinless Casana et al \cite{Casana:2017} Schwarzschild-like metric, obtained by putting $a_{\oplus} = 0$ in Eq.(41). However, the advancement effect, just like the Shapiro delay, is quite generically applicable to all types of metrics, whether approximate or exact. As stated after Eq.(18), the terms involving higher orders in $a$ are to be summarily ignored or at best accepted strictly in the Kerr limit, $\ell = 0$. The third reason is that the contribution of spin to time advancement could be too low under a specific condition: The magnitude of mass terms even as low as third order $R_{\text{S}}^{3}$ in Eq.(39) is \textit{larger} than the first order spin term ($aR_{\text{S}}$) if the impact parameter $b$ of the light trajectory satisfies the inequality, $b<\left( \frac{120-5\pi }{64}\right) \frac{R_{\text{S}}^{2}}{a} + \frac{R_{\text{S}}}{2}$, as shown by Wang and Lin \cite{Wang:2014} for the Kerr black hole. Under this condition, not much of spin contribution is actually lost while ignoring higher order terms in $a$. Unfortunately, this reasoning is inapplicable in our illustrative case of Earth-Moon system since Earth is not a black hole and so the constraint on $b$ cannot be realized in practice, since here $b = r_{B} >> R_{\text{S}} \sim 0.44$ cm. That's why the tables 1 and 2 show $\left\vert \Delta \tau _{(M_{\oplus }^{2})}\right\vert < \left\vert \Delta \tau _{(M_{\oplus }a_{\oplus })}\right\vert $ unlike for black holes. There is in general a larger question if any spinning black hole solution, say Kerr, can be legitimately applied to ordinary objects like Earth or Sun that give rise to the occurrence of multipoles. Therefore, the Earth-Moon system is to be regarded more as a pedagogical devise illustrating the Bumblebee effect on time advancement than a practical astrophysical situation involving black holes.

The effect of gravitational time advancement or negative time delay, first proposed in \cite{Bhadra:2010}, may be treated as yet another new test of general relativity complementary to the Shapiro time delay. It is new because it tells the other half of the known story of positive Shapiro time delay. Shapiro \cite{Shapiro:1964} obtained an \textit{excess} time over the time the signal would have taken to cover the two-way Euclidean distance $2\sqrt{r_{A}^{2}-r_{B}^{2}}$ valid in the absence of gravitating mass. On the other hand, time advancement means that the signal takes \textit{less} time than it would have taken covering the same Euclidean distance. However, three points need to be remembered: (1) As already cautioned in \cite{Bhadra:2010}, the negative Shapiro time delay or time advancement has nothing to do with backward time travel or warp drive etc. (2) We considered practical applications in Sec.4, in which light signals are assumed to travel back and forth in the spacetime of the centrally gravitating field of Earth only. (3) In either of the advancement/delay measurements, the difference in proper times $\left(\Delta \tau _{BAB}^{\text{total}}-\Delta \tau _{0}\right) $ in Eq.(42) and $\left( \Delta \tau _{ABA}^{\text{total}}-\Delta \tau _{0}\right)$ in Eq.(47) are recorded by a \textit{single} clock located respectively either at $B$ or at $A$ (Fig.1) after a round trip travel of signals.

We point out that, in addition to the slowing down of signals, there is another important underlying reason often omitted in the literature in the context of Shapiro delay. It is the \textit{universality} of gravitation manifested in the fact that clocks can measure only the unique proper time interval $\Delta \tau$ at any point in the gravity field and not the non-unique coordinate time interval $\Delta t$. \textit{Thus, the effects of delay and advancement result from a combination of two generic causes: the gravitational slowing down of signals in a gravity field and the universality of gravitation.} The known Shapiro experiment \cite{Shapiro:1971} is described by Eqs.(47-51) with the difference that now the signal propagates in the spacetime of Sun. Paraphrasing the experimental configuration in terms of our Fig.1, let us assume that a signal is sent from Earth $A$ to strong gravity of Sun $B$ (mass $M_{\odot }$, spin $a_{\odot }$) and reflected back to Earth $A$ that is located in the weak field of the Sun so that, after the two-way motion, the clock at $A$ measures $\Delta \tau _{ABA}^{\text{total}}=2\left( \sqrt{1-\frac{2M_{\odot }}{r_{A}}}\right) \Delta t.$ Since at the location of Earth, $\frac{2M_{\odot }}{r_{A}}<<1$, the proper time factor $\sqrt{1-\frac{2M_{\odot }}{r_{A}}}$ is legitimately neglected in the literature leading to $\Delta \tau _{ABA}^{\text{total}}\simeq 2\Delta t$, where $\Delta t$ is given by the Eq.(41). Despite this neglect, the value of $\Delta \tau _{ABA}^{\text{total}}$ after subtracting the Euclidean part $2\sqrt{r_{A}^{2}-r_{B}^{2}}$ can be easily understood to be confirmed in the paraphrased Shapiro time delay experiment and in our understanding, probably this is the reason why time advancement effect had not been noticed earlier. Note that Shapiro delay is a generic phenomenon valid beyond the original Earth-Sun-Mercury-Sun-Earth configuration and has played a central role in the determination of the mass of neutron star in a binary using the delay of signals passing by the strong field of pulsar and reaching the weak field of Earth \cite{Demorest:2010}.

In the above backdrop, let us summarize our results returning to the Earth-Moon configuration:

$\bullet $The signal travels from a point in strong gravity $B$ (Earth) to a point in weak gravity $A$ (Moon or any Satellite) and reflected back to the same point in strong gravity $B$. In this case, the factor $\sqrt{1-\frac{2M_{\oplus }}{r_{B}}}$ can no longer be ignored in Eq.(40). The result is a time advancement $\Delta \tau _{BAB}^{\text{adv}}:=\Delta \tau _{BAB}^{\text{total}}-\Delta \tau _{0}=\Delta \tau _{(M_{\oplus })}+\Delta \tau _{(M_{\oplus }^{2})}+\Delta \tau _{(M_{\oplus }a_{\oplus })}<0$ [from Eq.(42), Fig.2, for $\ell =0$] to be measured by the clock at $B$.

$\bullet $The novelty of the advancement effect is that the total measured distance between $B$ and $A$, viz., $c\Delta \tau _{BAB}^{\text{total}}/2$ becomes less than the zeroth order PPN Euclidean distance $c\Delta \tau _{0}/2$. This has the consequence that, to an observer sitting near the horizon of a black hole, the astrophysical objects will \textit{appear} arbitrarily closer depending on the observer's nearness to the horizon. The observable material universe would thus appear shrunken to him/her. This is not surprising as observations in general relativity depend on the location of the observer. However, arbitrary Euclidean distances are unmeasurable in gravity field.

$\bullet $There exists a critical radius $r=r_{c}$ (Fig.2) as a root of the Eq.(42): $\Delta \tau _{(M_{\oplus })}+\Delta \tau _{(M_{\oplus }^{2})}+\Delta \tau _{(M_{\oplus }a_{\oplus })}=0$. Addition of higher order terms to this equation from the expansion of Eq.(40) do not noticeably alter the position of $r_{c}$ since those terms are too small. Thus the clock at $B$ measures neither time delay nor advancement when the reflector $A$ is placed at $r=r_{c}$ indicating that the slowing down and universality approximately balance each other at the critical radius. If the reflector $A$ is placed at $r<r_{c}$, clock at $B$ measures usual Shapiro delay $\Delta \tau _{BAB}^{\text{delay}}$ and if the reflector $A$ is placed ar $r>r_{c}$, clock at $B$ measures time advancement $\Delta \tau _{BAB}^{\text{adv}}$.

$\bullet $We calculated the influence of Lorentz violating Bumblebee parameter $\ell $ on the advancement effect and its future measurement would enable us to put new constraints on $\ell$ in a manner suggested by Casana et al \cite{Casana:2017} using the experimental error in the original experiment of Shapiro time delay. Tables 1 and 2 show the order of magnitude of the expected time advancement values in the respective Earth-Moon and Earth-Satellite configurations. The listed values are only indicative and not exact as they are based on available data with varying degrees of accuracy. Nonetheless, we believe that the order of magnitudes predicted in the tables will not be altered. The measures of two clocks located at two points $A$ and $B$ differ in sign and magnitudes even though the Euclidean distance and the generic physics behind the effects are the same. This fact is explicitly demonstrated by the first row of the Table I: The clock at $B$ (strong gravity) at $\ell =0$ (say) is predicted to measure $\Delta \tau _{BAB}^{\text{adv}}=-1.460$ ns (or $-44.18$ cm), i.e., the Earth observer will see the Moon's distance to be apparently $c\Delta \tau _{BAB}^{\text{adv}}/2=22.09$ cm \textit{shorter }than the Earth-Moon Euclidean distance! Conversely, when a signal moves from Moon $A$ and reflected back to $A$ by Earth $B$, a clock at $A$ (weaker gravity) will measure Shapiro delay of magnitude $\Delta \tau _{ABA}^{\text{delay}}=+0.26$ ns ($+0.88$ cm), i.e., the Moon observer $A$ will conclude a distance $0.44$ cm \textit{longer} than the zeroth order PPN Euclidean Moon-Earth distance, since the Shapiro delay is always positive as shown in Fig.3. The readings are non-reciprocal because the end points of round trips, where clocks are located, are different.

$\bullet $The above non-reciprocity provides a nice reality check of the time advancement effect: As noted above, the clock at $A$ (Moon) measures a time delay $+0.26$ ns, while the clock at $B$ (Earth) measures a time advancement $-1.460$ ns. By arranging to collect information on their readings, one can get rid of the unmeasurable Euclidean distance, ending up with a difference of readings as $(\Delta \tau _{BAB}^{\text{total}}-\Delta\tau _{ABA}^{\text{total}})=(\Delta \tau _{BAB}^{\text{adv}}-\Delta \tau_{ABA}^{\text{delay}})=-1.20$ ns. \textit{This net negative value at once confirms the time advancement effect adopted in this paper.} Instead of Moon, one could employ Earth's Satellite and the same method could be applied though the required accuracy will be at the pico-second level. The effect of $\ell \neq 0$ is evident from Tables 1 and 2 and a very precise measurement of $\left( \Delta \tau _{BAB}^{\text{total}}-\Delta \tau _{ABA}^{\text{total}}\right) $, unavailable at the moment, can improve the constraint on the Bumblebee parameter $\ell $ beyond the one from the Cassini spacecraft constraint [see Eq.(57)]. However, the LSB correction ($\sim 10^{-4}$ ns) to the zeroth order Euclidean component due to the conical geometry of pure Bumblebee gravity, i.e., massless case (see Appendix) may be measurable in the near future. Some possibilities, though not exhaustive, for observational detection of the time advancement/delay effect have been outlined in Sec.5. For a useful discussion on precision tests of gravity, including Shapiro time delay, see \cite{Kramer:2014}.

It is understood that a direct actual measurement of the advancement effect, including the extended one [Eqs.(42-46) and (47-51)] developed in this paper by incorporating $a$ and $\ell $, is beset with many challenges arising out of the relative motion of Earth, Moon/Satellite, Solar influences, atmospheric disturbance etc not to mention the required level of accuracy of the measurements of all other concerned quantities. Such technical challenges notwithstanding, the reality of the novel effect of time advancement in general relativity under ideal conditions and the impact on it of spin $a$ and the LSB Bumblebee parameter $\ell $ cannot be missed.

\bmhead{Acknowledgments}

We thank an anonymous referee for many insightful comments that led to a considerable improvement of the paper. We are indebted to Gulnaz Kutlieva for her enlightening comments on different technical aspects of satellite communications relevant to the measurement of the time advancement effect.

\begin{appendices}

\section{Bumblebee corrections}\label{secA1}

Our idea is to separate pure Bumblebee corrections from the mass effects, so we put $M=0$ in the metric (19), and for simplicity consider the static case ($a=0$), which yields a conical spacetime%
\begin{equation}
d\tau^{2} = -dt^{2} + \left(1 + \ell \right) dr^{2} + r^{2}\left(d\theta^{2} + \sin^{2}\theta d\phi^{2}\right),
\end{equation}%
where $0 \leq \phi < 2\pi$. The Kretschmann scalar due to Bumblebee field is
\begin{equation}
R_{\mu\nu\alpha\beta} R^{\mu\nu\alpha\beta} = \frac{4\ell^{2}}{r^{4}(\ell + 1)^{2}}.
\end{equation}%
Assuming the signal to travel on the equatorial plane $\theta = \pi /2$, consider the 2-surface S:($R$, $\phi$) and using the method of Ford and Vilenkin \cite{Ford:1981}, the metric (A1) can be rewritten as a "flat spacetime" metric
\begin{equation}
d\tau ^{2}=-dt^{2}+dR^{2}+R^{2}d\phi ^{\prime 2},
\end{equation}%
where $b = \frac{1}{\sqrt{1+\ell }}$, $R=r\sqrt{1+\ell }$ and $0\leq \phi^{\prime} < 2\pi b$. Therefore the light ray moving tangentially along a straight line in the flat planar metric (A3) sweeps out an angle $\Delta\phi^{\prime}= \pi$ or%
\begin{equation}
\Delta \phi =\pi b^{-1}.
\end{equation}%
Hence the light ray undergoes a deflection of
\begin{equation}
\delta \phi =\pi b^{-1}-\pi =\pi \left( b^{-1}-1\right) =\pi \left( \sqrt{1+\ell }-1\right) \simeq \frac{\pi \ell }{2}.
\end{equation}%
Note that this is precisely the amount of correction to light bending by the LSB term, $\delta \phi _{\scriptsize{\textmd{LSB}}}$, obtained after elaborate calculations in \cite{Casana:2017}.

Similarly, the time delay in the "flat spacetime" (A3) is
\begin{equation}
\Delta t=\int \frac{RdR}{\sqrt{R^{2}-R_{0}^{2}}} = \sqrt{R^{2}-R_{0}^{2}},
\end{equation}%
where $R_{0}$ is the closest approach distance of the light ray or radar signal from the origin. Using $R=r\sqrt{1+\ell }$ and $R_{0}= r_{0}\sqrt{1+\ell }$, we find the Euclidean part of time delay for two-way motion to be
\begin{equation}
\Delta \tau _{0}=2\Delta t=2\left( \sqrt{1+\ell }\right) \left( \sqrt{r^{2}-r_{0}^{2}}\right) =\delta \tau _{0}+\delta \tau _{\scriptsize{\textmd{LSB}}}^{\scriptsize{\textmd{Eucl}}}.
\end{equation}%
This is precisely Eq.(52), which is what we wanted to show. The correction to the Euclidean part of time delay for two-way motion in the original ($t,r$) coordinates, restoring $c$, then is%
\begin{equation}
\delta \tau _{\scriptsize{\textmd{LSB}}}^{\scriptsize{\textmd{Euc}}}\simeq \frac{\ell }{c}\sqrt{r_{A}^{2}-r_{B}^{2}},
\end{equation}%
confirming the LSB correction obtained after elaborate calculations in \cite{Casana:2017}.

\end{appendices}


\begin{thebibliography}{}
\bibitem{OBrian:2004}
O'Brian, T., Diddams, S.: What is the fastest event (shortest time duration) that can be measured with today's technology and how is this done?
Scientific American (2004).
\end{thebibliography}
\end{document}